\documentstyle[twoside,epsfig,fleqn,espcrc2]{article}
\def\aprle{\buildrel < \over {_{\sim}}}

\newcommand{\ndt}{\noindent}
\newcommand{\ov}{\overline}
\newcommand{\AmS}{{\protect\the\textfont2
  A\kern-.1667em\lower.5ex\hbox{M}\kern-.125emS}} 

\title{Highlights in Neutrino and Astroparticle Physics}

\author{G. Giacomelli\address{Dipartimento di Fisica dell'Universit\`a di
        Bologna and INFN,
        Sezione di Bologna,\\
        viale Berti Pichat 6/2, I-40127 Bologna, Italy \\
        e-mail: giacomelli@bo.infn.it}}

\begin{document}
\begin{abstract}
{\bf Closing lecture of the 1999 S. Miniato Workshop, {\it Neutrino and Astroparticle
Physics},\\
S.Miniato, Italy, 17-21 May 99.}\\

The main highlights from the papers presented at this workshop are briefly
 reviewed and discussed in a general context.
\end{abstract}

\maketitle

\section{INTRODUCTION}
This was a timely workshop, because of the many new important experimental
results on neutrino physics and astrophysics, and the renewed interest in
 astroparticle physics, both experimentally and theoretically.

The experimental evidences on neutrino oscillations are mounting, and many new
 experiments are being planned to definitely prove these indications and   
accurately measure the neutrino oscillation parameters.

The astrophysical $\gamma$-ray bursts seem to be a dominant phenomenon in our 
universe.
New, larger and more sophisticated detectors are planned to study the higher 
energy $\gamma$-rays. 

Neutrino astrophysics started in the 1960's with the first detection of
 electron neutrinos from the sun, and was in a sense reborn in 1987 with
  the detection of electron antineutrinos from Supernova SN87A. Many detectors
 are now ready to study these two phenomena. High energy 
muon neutrino detectors of large volumes are entering the scene,
 possibly opening the field of
 high energy muon neutrino astronomy.

Many search experiments are trying to detect possible components of the Dark
 Matter (DM), or search for new particles predicted by the Standard
Model (SM) of particle physics, and by theories which go beyond the SM.

The study of the highest energy cosmic rays is another field of interest,
 in particular for determinig the mechanisms responsible for their 
acceleration. Very large area detectors, above and below ground, are needed to
 study this field.

A Large effort is being made to develop instrumentation for the detection of
 gravitational waves, which should reveal some of the most violent phenomena
 occurring in the cosmos. The detectors include supercooled antennas at 0.1 K
 and very long
interferometers (few km).

In this summary of the workshop I shall recall many of the papers presented,
 with special emphasis on neutrino oscillations and on higher energy 
phenomena; I shall not be able to cover in detail all subjects, nor quote
 all
results. I apologize for this impossibility and for possible omissions.

During the workshop we were informed of the sudden death of Bianca Monteleoni
 Conforto, a colleague and a collaborator. I dedicate these notes to her memory.

\section{NEUTRINO PHYSICS}

Most of the interest in this field concentrated on experimental
 results relevant to neutrino
 oscillations.    

\subsection{Atmospheric neutrinos}
   The interest in atmospheric neutrinos has 
grown in the last year, after the Neutrino '98 Conference in Takayama, 
Japan. New, higher statistics 
data have been presented by the Soudan 2 \cite{soudan},
MACRO \cite{ronga98}, and SuperKamiokande (SK) \cite{sk98} \cite{kajita}
 collaborations. 
The measured flux of muons induced by atmospheric 
$\nu_\mu$ shows
a reduction with respect to the expectation; the reduction depends on the
neutrino energy and direction. For $\nu_e$ induced  electrons 
there is no deviation from the prediction. 
The three experiments can
explain the $\nu_\mu$ reduction 
in terms of $\nu_\mu \rightarrow \nu_\tau$ neutrino oscillations, 
with maximum mixing and $\Delta m^2$ values of few times
$10^{-3}\ eV^2$.

 In the simplest scenario of two flavor oscillations,
the survival probability of a pure $\nu_\mu$ beam is

\begin{equation}
P(\nu_\mu \rightarrow \nu_\mu) = 
1- sin^2 2\theta\ sin^2 ( { {1.27 \Delta m^2 \cdot L}\over {E_\nu}})
\end{equation}
 where $\Delta m^2 = m^2_2 - m^2_1$ is the mass
 difference of the two neutrino mass states,
 $\theta$ is
the mixing angle, $E_\nu$ is
the neutrino energy and $L$ is the path length from the $\nu_\mu$
production point to the detector. For atmospheric neutrinos
 $L$ can be estimated through  
the neutrino arrival direction $\Theta$. For 
upgoing neutrinos, as the zenith angle $\Theta$ changes, one has
$L \sim 2R_\oplus\cdot cos\Theta$ ($R_\oplus$ is the Earth radius), while
$L$ is only few tens of kilometers for vertical downgoing neutrinos.

\begin{figure}[htb]
\vspace{-0.3cm}
\begin{center}
\mbox{
\epsfig{file=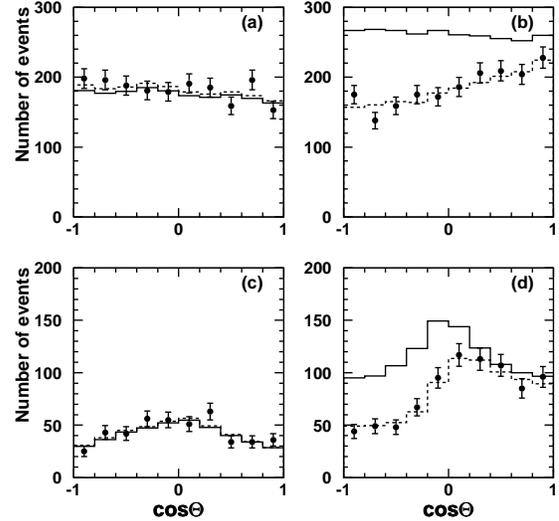,height=7cm}}
\end{center}
\vspace{-0.8cm}
\caption{\small SK data. Zenith angle distributions of sub-GeV 
(a) $e$-like events
 and (b) $\mu$-like, (c) multi-GeV $e$-like,
 (d) multi GeV $\mu$-like data. Upward going particles have $cos \theta
 \aprle 0$. The black points are the data, the solid lines are the
 Monte Carlo expectations for no-oscillations; the dashed lines 
are the MC predictions for $\nu_\mu \rightarrow \nu_\tau $ oscillations
with $\Delta m^2 = 3.5 ~10^{-3} e V^2$ and $sin^2 2 \theta = 1$ [3][4].}
\label{fig:skdata1}
\vspace{-0.8cm}
\end{figure}

Atmospheric neutrinos are detected in the SuperKamiokande 
(SK) water \^Cerenkov detector via their interactions with p and $^{16}O$
nuclei in the $22500~m^3$ water fiducial volume ($50.000~ m^3$ total volume).
 Three different classes of events are 
defined (with increasing average energy of the parent neutrino): 
fully contained events (FC), partially contained events (PC)
and upward-going muons. FC events are further subdivided into sub-GeV and
 multi-GeV.
Electrons are identified in the FC
sample. The zenith angle distribution for e-like sub-GeV and multi-GeV events
 are in reasonable agreement with the predictions assuming no-oscillations.
Instead the $\mu$-like events deviate considerably from the prediction, see 
Fig. \ref{fig:skdata1}.
The ratio of the measured numbers of muons 
to electrons normalized to the respective
Monte Carlo predictions is affected by a smaller systematic error, and it
 enhances the anomaly \cite{kajita}. 

\begin{figure}[ht]
\vspace{-3cm}
\begin{center}
\mbox{
\hspace{-0.8cm}
\epsfig{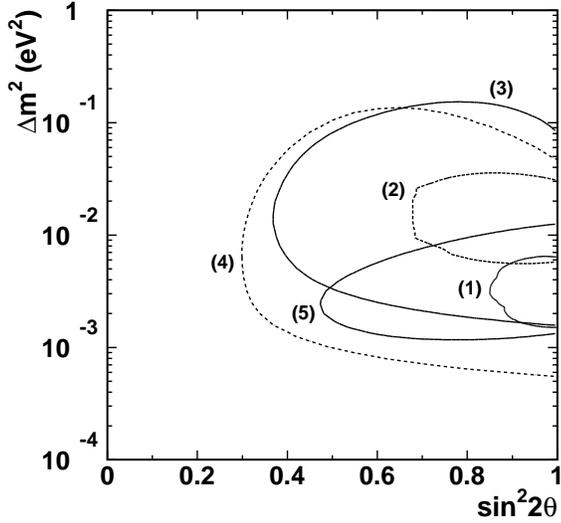}}
\end{center} 
\vspace{-3cm}
\caption{\small The $90 \%$ C.L. contours in the $sin^2 2 \theta$ and
 $\Delta m^2$
 plane for $\nu_\mu \rightarrow \nu_\tau$ oscillations for the Kamiokande and 
 SuperKamiokande data [4]. (1) and (2) refer to contained events from SK
 and Kamiokande, respectively. (3) and (4) refer to upward through-going muons
 from Super-K. and K. (5) shows the region obtained by the
 (stopping/through-going) ratio
 of upward going muons from SK.}
\label{fig:c.l.}
\vspace{-0.8cm}
\end{figure}

Assuming two flavor oscillations Fig.\ref{fig:c.l.} 
shows the $90 \%$ C.L. contours
delimiting the accepted regions by Kamiokande and SK. The SK data favour
$\nu_\mu \rightarrow \nu_\tau$  neutrino oscillations with 
$ \Delta m^2 = (1.5-6)10^{-3} e V^2$ and $sin^2 2 \theta > 0.9$. For more
 details on further data on upthroughgoing muons and stopping muons,
see the paper presented by Kajita at this workshop \cite{kajita}, and 
\cite{sk98}. SK obtains also indications for an east-west asymmetry, which may
be considered a confirmation of the flux calculations and of the experimental 
methods \cite{kajita}.

\begin{figure}[ht]
\begin{center}
\mbox{
\hspace{-1.6cm}
\epsfig{file=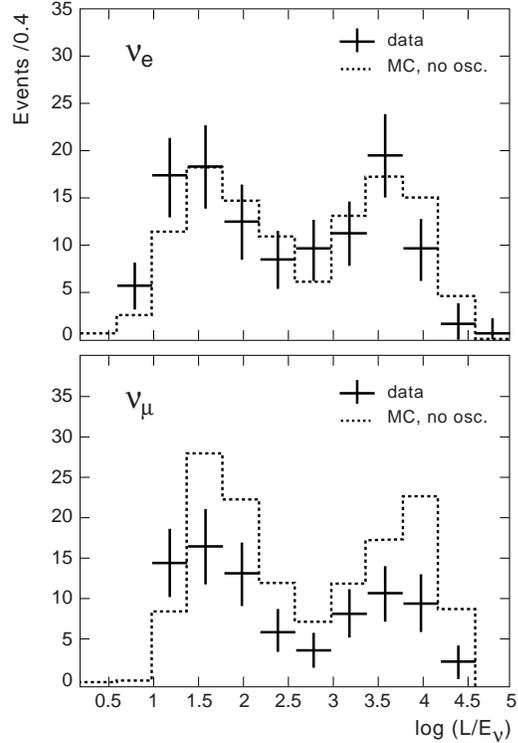,height=10cm}}
\end{center}
\vspace{-0.8cm}
\caption {\small Soudan 2 results on the number of observed $\nu_e$ (top)
 and $\nu_\mu$ (bottom) events as function
 of $L/E_\nu$.  Only statistical errors are shown;
 the dashed lines are the MC predictions assuming no-oscillations. Note the 
lack of $\nu_\mu$ events at  $Log_{10}$ (L/E) $\simeq 1.8$ and $3.9$.}
\label{fig:soudan2}
\vspace{-0.8cm}
\end{figure}

The Soudan 2 results support the oscillation hypothesis
by measuring atmospheric $\nu_\mu$ and $ \nu_e$ interactions at
low energies, below $1$ GeV \cite{soudan}. A different detection technique
(drift chamber calorimeter) is used in this case; the total mass of the 
detector is about $1~ kt$ and the total exposure is $4.6~ kt~ y$. 
Fig. \ref{fig:soudan2} shows, for the high-resolution contained data, vs Log 
(L/E$_\nu$) relative to the no-oscillation expectations. Still with low
 statistical significance, the data agree with a reduction of $\nu_\mu$ events
compared to expectations, while the $\nu_e$ events agree with expectations.

\begin{figure}[htb]
\centerline{
\epsfig{figure=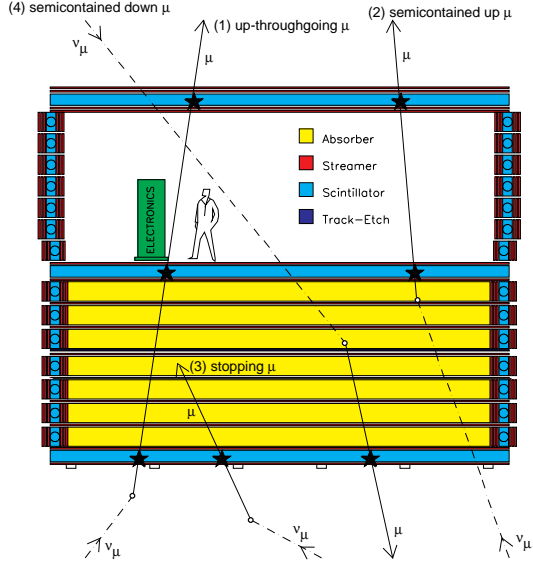,height=7.5cm} }
\vspace{-0.8cm}
\caption {\small Sketch of different event topologies 
induced by muon neutrino interactions in or around MACRO. 
The stars represent the liquid scintillator hits.}
\label{fig:topo}
\vspace{-0.8cm}
\end{figure}

The MACRO detector is a box of
 $76.6~m~\times~12~m~\times~9.3~m$ located at
 the Gran Sasso Lab.; the detection elements are planes of limited 
streamer tubes for tracking and liquid scintillation counters for timing.
The lower half of the detector is filled
with trays of absorbers alternating with streamer tube
planes, while the upper part is open. 
The angular resolution for muons achieved by the streamer tubes 
is $< 1^\circ$. The
time resolution of the scintillators is $0.5\ ns$.
Fig. \ref{fig:topo} shows the measured topologies.

The up throughgoing muons come from $\nu_\mu$
 with $ \overline E_{\nu_\mu} \simeq 80~ GeV$ interacting in the rock
 below the detector; their flight direction is determined by
time-of-flight ({\it t.o.f.}).
 $\nu_\mu$ with $ \overline E_{\nu_\mu} \simeq 4~GeV$ interact
 inside the lower apparatus;
yielding upgoing muons ({\bf IU}). The partially contained downgoing events ({\bf ID}) 
and upward going stopping muons  ({\bf UGS}) are 
identified via topological constraints.

Monte Carlo simulations play a crucial role in atmospheric neutrino studies.
Macro used the Bartol neutrino flux
\cite{agrawal} and the deep inelastic scattering (DIS) parton
distribution of ref. \cite{gluck} for the neutrino cross-sections.
The propagation of muons
is done using  the energy
loss calculations of ref. \cite{lohmann} in standard rock.
The systematic uncertainty on the expected flux of muons
 is $\pm17\%$;
this 
is a scale factor, that changes little the shape of the angular
distribution.

\begin{figure} [htb]
\centerline{
\epsfig{file=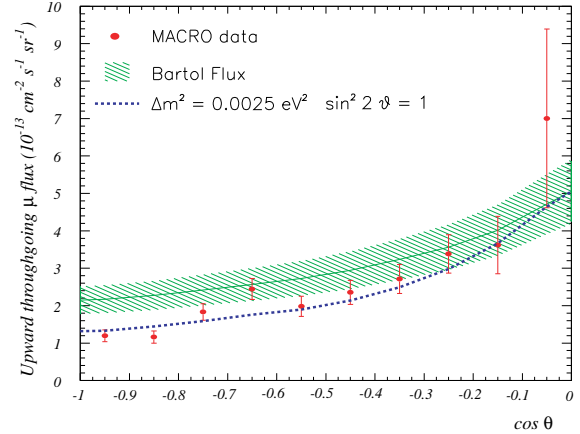,height=5.8cm} }
\vspace{-0.8cm}
\caption{\small Measured flux (points) of upward throughgoing muon $vs.$
 the cosine of
zenith angle $\Theta$. The solid line is the prediction with no oscillations;
 the 17\% scale
uncertainty is shown as the dashed region, the error on the shape is  almost
 negligible.
The dashed line shows the prediction assuming two-flavor neutrino
oscillations.}
\label{fig:flux}
\vspace{-0.8cm}
\end{figure}

The ratio of the observed to expected number of upthroughgoing muons
is $0.74\pm0.031_{stat} \pm 0.044_{sys} \pm 0.12_{theo}$.
Fig. \ref{fig:flux}  shows their zenith angle distribution 
compared to Monte Carlo expectation without neutrino oscillations (solid line);
the dashed line is the fit to the data assuming $\nu_\mu \rightarrow \nu_\tau$
oscillations.
The reduction in the detected number of events
and the deformation of the zenith angle distribution may be due
to $\nu_\mu$ disappearance:
fewer events are expected near the vertical 
$(cos\Theta=-1)$ than near the horizontal $(cos\Theta=0)$,
due to the longer path length of neutrinos from production to observation.  
The maximum of the $\chi^2$ probability corresponds to
$\Delta m^2 =2.5\times 10^{-3} eV^2$ and maximum mixing.
The confidence region at the 90\% C.L. 
in $(\sin^2 2 \theta , \Delta m^2)$ for
$\nu_{\mu}  \rightarrow\nu_{\tau}$ oscillations agrees and is somewhat
 larger than that
of SK \cite{ronga98}. Notice the possible excess of events at
$cos\Theta \sim -0.65$ (also the up-throughgoing muons of SK have a similar
 hint); it is consistent with 
a statistical fluctuation, but it could be a hint for a more complex scenario.

\begin{figure}[htb]
\vspace{0cm}
\centerline{
\epsfig{figure=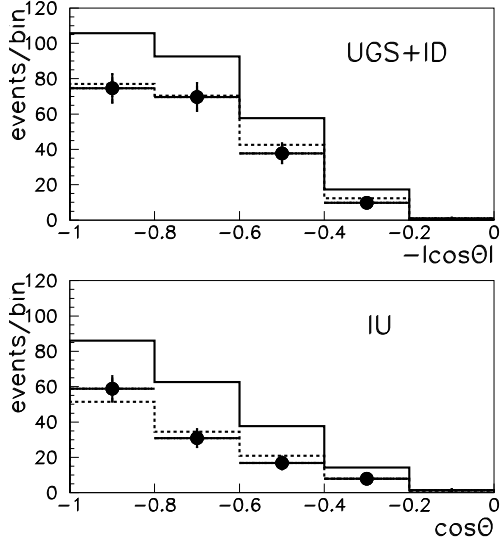,height=7.2cm}}
\vspace{-0.8cm}
\caption {\small MACRO Zenith angle  distributions for
(ID+UGS) and IU events.
The data (black points with error bars)
are compared with the Monte Carlo
expectations assuming no oscillations (full lines) and two-flavour
oscillations (dashed lines) using maximum mixing and
$\Delta m^2=2.5 \times 10^{-3}\ eV^2$.}
\label{fig:lebo}
\vspace{-0.8cm}
\end{figure}

Fig. \ref{fig:lebo}  shows the zenith angle distribution of 
the semicontained (IU) and upstopping muons plus partially contained downgoing
 muons (UGS + ID).
The data are within errors consistent with a constant
deficit  with respect to the MC expectations.
The ratios of the number of observed to expected events are
$R_{ID+UGS} = ({{Data}\over {MC}})_{ID+UGS} \simeq
0.71$ and
$R_{IU} \simeq 0.57$.
The theoretical and systematic 
errors are largely reduced (from 25\% to about 5\%) 
if the ratio of ratios is considered, ${\cal R}=R_{IU}/ R_{ID+UGS}=
0.80\pm 0.09_{stat}$; for no oscillations one expects $R=1$. 
 The reductions are consistent with
$\nu_\mu \rightarrow \nu_\tau$ oscillations with maximum mixing and
$\Delta m^2 \sim 10^{-3} \div  10^{-2}\ eV^2$.

Several theoretical papers tried to interpret the atmospheric neutrino data in
 terms of oscillations among three neutrino types; the differences with the
 simpler $\nu _\mu - \nu_\tau$ possibility are small \cite{fogli}. Other
authors considered $\nu_\mu \rightarrow \nu_{sterile}$,
 which is slightly disfavoured
 by the data. Others include $\nu_e , \nu_\mu , \nu_\tau , \nu_{sterile}$.

G. Battistoni \cite{battistoni} discussed the effects of the approximations
used in the Monte Carlo predictions. Present MCs use the collinear approach,
 which cannot be a good approximation at low energies. But 3-D
 effects are smeared out 
because of Fermi motion of the nucleons in nuclei. Uncertainties remain in the
 knowledge of primary cosmic ray spectra, secondary particle production and
 neutrino cross sections. Hopefully measurements of muons in the
 atmosphere \cite{circella} 
could improve the predictions, though the sub-GeV range remains
 problematic. 

Improved atmospheric neutrino detectors are under 
discussion \cite{terranova}. 
   
\subsection{Solar neutrinos}
Solar $\nu_e$ come from a chain of nuclear reactions and decays in the centre 
of the sun. The three important components of the spectrum are: i) the 
energetic neutrinos from $^8B$ decay; ii) the monoenergetic neutrinos 
from $^{7}$Be+$e^-\rightarrow$ $^7$Li+$\nu_e$ 
$(\mbox {E}_{\nu_e}=0.862$ MeV) and iii) the low--energy part, the $pp$
 neutrinos $(\mbox {E}_{\nu_e} \leq  0.41$ MeV) (most abundant) .

Experimental measurements of solar neutrinos have been performed by five 
experiments using three different reactions \cite{bellini}. \par

The first measurement of solar neutrinos used a radiochemical 
method via inverse $\beta$ decay,
~~$ \nu_e+^{37}$Cl$\rightarrow ^{37}$Ar$+e^-$~~, 
which has a neutrino energy threshold  ~$\mbox{ E}_{\nu th}=814$ keV.
The experiment, sensitive to $^7 B e$ and $^8 B$ neutrinos 
yields a flux smaller than that predicted by the Standard 
Solar Model (SSM):
$ (\phi_{7_{Be}}+~ \phi_{8_B})(\mbox{Cl})<(\phi_{7_{Be}}+~\phi_{8_B})(\mbox{SSM})$.

The second measurement was performed in the Kamiokande water \^Cerenkov
 de\-tec\-tor using the reaction
~~$ \nu_e+e^-\rightarrow \nu_e+e^-$.  
They apply a cut at E$_{\nu th} \simeq 7~ MeV$ and are thus
 sensitive only 
to $^8B$ neutrinos. The angular distribution is peaked in the direction of the 
sun and therefore confirms that the detected neutrinos come from the sun. They
obtained a ratio $expected/measured=0.417\pm 0.069$.
Combining this with the chlorine result, one has  
a discrepancy expressed as  
$ (\phi_{7_{Be}}+~ \phi_{8_B})(\mbox{Cl})<\phi_{8_B}(\mbox{Ka})$.

A third reaction is studied by radiochemical methods using ~$^{71}Ga$ in 
metallic (SAGE) and in a hydrochloric water solution (GALLEX, GNO):
~~$ \nu_e+ ^{71}$Ga$\rightarrow ^{71}$Ge$+e^- $~~, 
 which has a threshold at E$_{\nu_e}$=233 keV. Thus one may measure the 
neutrinos coming from the $pp$ reactions, proving that the sun is a $pp$  
nuclear fusion plant. 
The experiments yield values smaller than the SSM prediction; these low
 values and the  comparison with the preceeding measurements, lead to
 $ \phi_{7_{Be}}(meas)<\phi_{7_{Be}}(\mbox{SSM})$, which seems to be the main
 problem.

\begin{figure}[ht]
\vspace{-0.5cm}
\begin{center}
\mbox{
\hspace{-0.4cm}
\epsfig{file=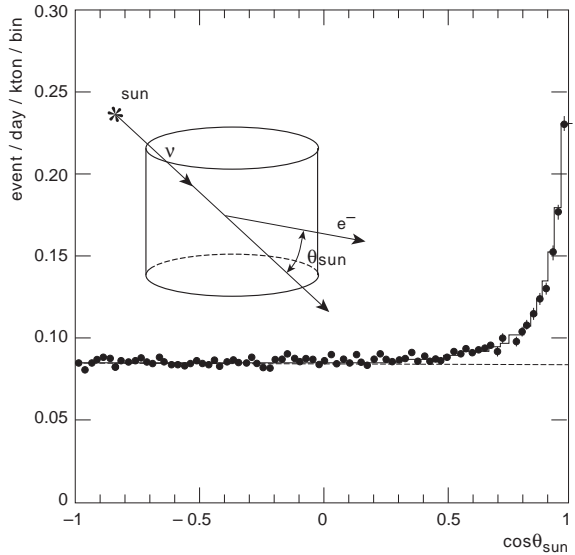, height=8cm}}
\end{center}
\vspace{-1cm}                                                                      
\caption{\small Data from SuperKamiokande on the arrival direction of solar
 neutrinos
measured via the reaction $\nu_e e^- \rightarrow \nu_e e^-$. Notice the peak at 
$cos \theta = 1$, towards the sun.}
\label{fig:data}
\vspace{-0.5cm}
\end{figure}

Superkamiokande presented at this workshop new results on solar neutrinos 
which further confirm the above statements. The data also confirm that the
 solar
$\nu_ e$ come from the sun \cite{kajita}, see Fig. \ref{fig:data}. A 
day-night effect might have been observed by SK: this would be expected for the
MSW effect. Also a seasonal variation due to the eccentricity of the earth
 orbit might have been observed.

The lack of observed Be solar neutrinos seems at present to be the essence  
of the solar neutrino problem.  
This could be due to a faulty experiment. Assuming 
that the experiments are correct, physicists looked for an 
astrophysical solution and at neutrino oscillations. Improvements
have been made in the knowledge of the sun interior \cite{bellini}
 and it seems that one cannot
 explain the deficits. One is therefore left with the possibility of neutrino
 oscillations.

\begin{figure}[htb]
\vspace{-0.3cm}
\begin{center}
\mbox{
\hspace{-1cm}
\epsfig{file=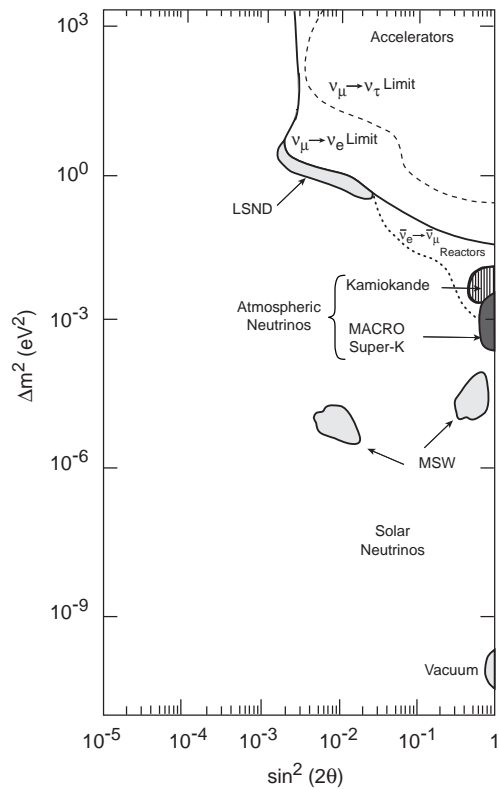,height=10.4cm}}
\end{center}
\vspace{-1cm}
\caption{\small Compilation of excluded regions by short 
baseline accelerator and reactor experiments, and of allowed regions by the 
LSND accelerator experiment [13], by the atmospheric $\nu_\mu$ experiments and
 by the solar neutrino experiments, assuming MSW matter oscillations and vacuum
oscillations. More updated compilations are given in [4], [12]-[18].}
\label{fig:regions}
\vspace{-0.8cm}
\end{figure}

 Possible solutions of the problem assuming neutrino oscillations 
in vacuum and possible 
solutions assuming 
neutrino oscillations in solar matter (the MSW effect) are indicated in
the compilation of  Fig. \ref{fig:regions}; more up to date compilations have
 been presented in  \cite{terranova}-\cite{stancu}; the MSW low-mixing solution
seems to be disfavoured.

 Most solar neutrino experiments are relatively low--rate 
experiments. Superkamiokande and future experiments  have higher
 rates and have more specific aims.  
The SNO detector in Sudbury, Canada is starting to take data with neutrino
 interactions in $D_2O$. At Gran Sasso the Borexino experiment plans to
 detect $^7Be$ neutrinos via the 
reaction $\nu_ee^-\rightarrow \nu_ee^-$ in liquid scintillators \cite{bellini};
 ICARUS should
detect $\nu_e$ interactions in an $^{40}Ar$ TPC \cite{kignoli}. 
 There are discussions about the possible use of Li I (Eu) scintillation
 counters.

\subsection{Accelerator and reactor experiments}

We have heard numerous reports about short baseline accelerator 
experiments: LSND,
KARMEN, NOMAD, CHORUS and the future MINIBOONE. LSND gave a possible signal
for neutrino oscillations for $\Delta m^2 \sim 1~ e V^2$ and
 $10^{-3} \aprle sin^2 2 \theta <$ few $10^{-2}$, see Fig. \ref{fig:regions}
\cite{stancu}. The other experiments gave limits, which are globally 
summarized in Fig. \ref{fig:regions}.
Several technical improvements were made by these experiments; I shall only
 recall the revival of the emulsion technique for neutrino physics, 
specifically for $ \nu_\tau $ detection, because of the exceptional  space
 resolution
 ($ \sim 1~ \mu m $) of the technique \cite{kawanura}.

The results from the Chooz and Palo Verde reactor experiments exclude
 $ \nu_e \rightarrow \nu_\mu $ oscillations for 
$\Delta m^2 > 10^{-3} e V^2$ and $2 \times  10^{-2} < sin^2~ 2\theta < 1$,
not shown properly in Fig. \ref{fig:regions}.

\subsection{Long baseline experiments}

Several long baseline experiments have been proposed; they will cover the 
region $\Delta m^2 >  10^{-3}~ eV^2$ and $sin^2 2 \theta >  10^{-2}$
 \cite{harris}
 \cite{cocco}.\\
- K2K: from KEK to SuperKamiokande ($230~ km$) is starting to take data;
 they also
have a near detector \cite{kajita}.\\
- MINOS: from the Fermilab main injector to the Soudan mine ($730~ km$)
 is under construction; it is a $6~ kton$ calorimeter detector; they will
 also have a near detector \cite{harris}.\\ 
- KAMLAND: $\overline \nu_e$ from nuclear reactors will be detected in the
 Kamiokande
 mine with a liquid
 scintillator detector \cite{kamland}.\\
- ICARUS $\rightarrow$ now ICANOE: from CERN-SPS to Gran Sasso ($730~ km$)
 \cite{kignoli}; ICARUS will be an "electronic" bubble chamber; NOE a tracking 
calorimeter detector.\\
- OPERA: from CERN to Gran Sasso \cite{cocco}; it is basically a (large) emulsion
detector.\\

\subsection{Direct measurement of the ${\bf \overline \nu_e}$ mass}
Tritium decay, $t \rightarrow  ^3{He} + e^- + \ov\nu_e$ with
  Q $ = 18.6~ keV,~ t_{1/2} = 12.3~ y$, has been used by many groups to obtain
increasingly better limits on the $\ov\nu_e$ mass ($ \simeq \ov\nu_1$ mass 
if there is 
small mixing). The techniques and the calculations have been constantly
 improved. The latest results with the improved Mainz set up give a $95 \%$
 C.L. upper limit $ m_{\overline\nu e} \aprle 2.8~ e V$ \cite{weinheimer}.

\subsection{Neutrinoless double beta decay}
For even-even nuclei the chain decays

$(A,Z) \rightarrow (A,Z+1)+e^- + \overline \nu_e ~,\\
(A,Z+1) \rightarrow (A,Z+2)+e^- + \overline \nu_e $~~~~~~~~~~~~~~~~~~(2a)
\ndt are forbidden by energy conservation; the decay may be possible in a 
single step: \par
$(A,Z) \rightarrow (A,Z+2)+2e^- +2 \overline \nu_e,$~~~~~~~~~~~(2b)\\
$(A,Z) \rightarrow (A,Z+2)+2e^- +x,$~~~~~~~~~~~~~(2c)\\
$(A,Z) \rightarrow (A,Z+2)+2e^- $~~~~~~~~~~~~~~~~~~(2d)

The neutrinoless double beta decay, Eq. (2d), is forbidden by lepton number 
conservation; it would be allowed if $\nu_e$ and $\overline \nu_e$ were 
identical and if they had a non-zero mass. The energy spectrum for the sum of
 the energies of the two electrons, $E = E_1 + E_2$, is different for each of
 the three cases: a line for (2d), a continuum peaked at low $E$ for
 (2b) and a continuum peaked at higher $E$ for (2c).

Most of the direct searches for neutrinoless double beta decays use materials
which act both as source and detector, such as $^{76}Ge \rightarrow
 ^{76}Se + 2~e^{-2}$ \cite{fiorini}. Germanium detectors ranging from 1
 to $7~ kg$ have been used.
Normal germanium contains $15\%~ ^{76}Ge$. The Heidelberg-Moscow Collaboration
 uses several kilograms of enriched germanium containing $85 \% ~^{76}Ge$. 
From an exposure of $24~ kg~ y$ they quote $t_{1/2} > 6 \times 10^{25}y ~
(90 \% CL)$,
which in certain models corresponds to  $m_{\nu_e} <0.2~ eV$. Some groups use
visual detectors, separating the spatial detection of the two electrons. The 
double beta decay $^{136}Xe \rightarrow ^{136}Ba + 2e^-$ has a favorable 
transition energy of $2479~ keV$.

Considerable work is going on in the development of cryogenic detectors for
double beta decays and for dark matter searches \cite{fiorini}.
 At low temperature the heat
 capacity is very small, and a small energy deposition implies a relatively
 large increase in temperature. Four cryogenic crystals of Te $O_2$, each 
about $340~ g$, were used by the Milano-Gran Sasso collaboration to study the
  double beta decay of $^{130}Te$. Four sapphire detectors, each of $262~ g$,
are used by the CRESST experiment in a search for dark matter WIMPs.   

\section{NEUTRINO ASTROPHYSICS}

\subsection{Neutrino Astronomy}

One of the main interests in neutrino astronomy is connected with the great 
penetrating power of neutrinos, which allows us to look directly at their 
sources. The universe is filled with fossil low-energy neutrinos from the 
Big Bang. 
 Low--energy neutrinos of $\sim 1$ MeV come continuously from the 
interior of stars like the sun; slightly--higher--energy neutrinos $(\sim 12$ 
MeV) come in bursts from supernovae explosions. High--energy 
neutrinos ($>1$ GeV), may come from non--thermal point sources.
 Neutrinos of $>1$ GeV may also come 
from the sun and the earth, where annihilations of 
 WIMPs could take place.

\subsubsection{Neutrinos from stellar gravitational collapses}

Massive stars, $m>6~m_{\odot}$,  evolve as 
increasingly heavier nuclei are produced and then burnt at their centres in a 
chain of thermonuclear processes, ultimately leading to the formation of a 
core composed of iron and nickel. When the core mass exceeds the Chandrasekar
 limit,  
the core implodes in a time slightly longer than the freefall 
time and leads to the formation of a neutron star. 
The  energy released during a stellar collapse is at least the 
gravitational binding energy of the residual neutron star, $E\simeq 3\times 
10^{53}~(m/m_{\odot}^2)$ (10~$km$/R) $ergs$,  $\simeq 10^{53}~ ergs$ 
$\simeq$ 0.1 
$m_{\odot}$, mostly in the form of neutrinos with
 $\langle E_{\nu}\rangle\simeq ~  12~ MeV$. 
About $4\times 10^{57}$ neutrinos of each species are emitted. Three 
stages of neutrino emission may be identified.\par

All types of neutrinos may be detected via  neutral current interactions 
with electrons, $\nu_e e^-\rightarrow \nu_e e^- , ~\ov \nu_e e^-\rightarrow 
\ov\nu_e e^-$, etc, with a cross section $\sigma=1.7\times 10^{-44}$ E$_{\nu}$ 
($MeV cm^2$). The dominant  
reaction,  $\ov\nu_e p\rightarrow n e^+$, with $\sigma=7.5\times 
10^{-44}~E^2_{\nu}$ ($MeV^2 cm^2$), is energetically possible only on free 
protons, as in H$_2$O and in C$_n$ H$_{2n+2}$ detectors. The positron produced 
annihilates immediately, $e^+e^-\rightarrow 2\gamma$, whilst the neutron is 
moderated and captured after a mean time of about 180 $\mu$s ($n p\rightarrow 
d\gamma$, with $E_{\gamma}\sim 2.2$ MeV). The SNO  
detector with $D_2O$ will also 
detect $\nu_e n\rightarrow p e^-$. Because of the dependence of the cross 
section on neutrino energy, the average $e^+$ energy is about 2 MeV 
larger than the average $\overline \nu_e$ energy.

 Only Supernova SN1987A in the Large 
Magellanic Cloud was observed with neutrinos. No other burst of
 supernova neutrinos has been 
detected. Present and future neutrino 
detectors, will only be able 
to observe 
 galactic  supernovae. 
 An optimistic estimate of the rate of type-II Supernovae in our galaxy is 
one every 10--20 years. Several detectors are kept alive all the time
 and a worldwide supernova watch is in operation.

\subsubsection{High-energy neutrino astronomy}

High-energy muon neutrinos can be detected via their charged-current 
interactions inside a detector or 
in the rock surrounding the detector leading to upgoing muons. 
Upward--going muons can be seen directly in \^Cerenkov 
detectors and can be separated by time--of--flight from downward--going muons in 
scintillators. At very high energies the $\nu_\mu -\mu$ angle is small and the 
effective target may be large. In order to observe celestial "point"
 sources of high--energy  
neutrinos one should plot for each muon its declination versus
 right ascension. A celestial
source would reveal itself as an excess of events (in a certain direction) 
above the atmospheric neutrino background. Now, people are also looking at
time coincidences with $\gamma$ -ray bursts.\par
Several underground experiments performed searches for astrophysical sources
of $\nu_\mu$ , with negative results. 
In order to establish a flux limit for a specific source one may consider an
 error circle
corresponding to the resolution of the detector in that direction 
( $\simeq 3^{\circ}$ for tracking detectors, considerably
more 
 for  H$_2$O \^Cerenkov detectors),  
determine the number of events in that circle, and subtract the corresponding
number of events expected from atmospheric neutrinos. 
MACRO, with about 1000 muon events, quotes limits at
 the $10^{-14} cm^{-2} s^{-1}$ level \cite{ambrosio}.

{\bf Neutrino telescopes.} Much 
larger detectors, the so called Neutrino Telescopes, will 
be required to really attack the field of $\nu_\mu$ astronomy.
 Prototypes of neutrino telescopes 
may be considered the \^Cerenkov detectors NESTOR, ANTARES and NEMO
 under deep sea 
water, Baikal under lake water and AMANDA under ice at the South Pole 
\cite{telescopes}. The 
final detector will be around $1~km^3$ of water or ice.

\subsection{Searches for WIMPs}

\begin{figure}[ht]
\vspace{-1cm}
\begin{center}
\mbox{
\hspace{-0.2cm}
\epsfig{file=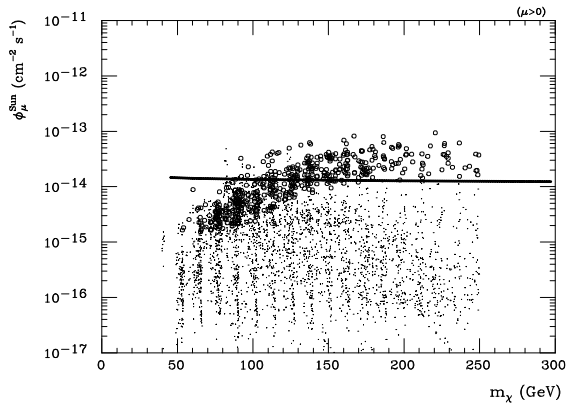,height=5.3cm}}
\end{center}
\vspace{-1cm}
\caption {\small  Upward-going muon flux vs neutralino mass $m_\chi$ from 
the Earth.
Each dot is obtained varying model parameters, leaving $\mu > 0$. Solid line:
MACRO flux limit ($90\%$ C.L.); 
the  limit for the  no-oscillation
hypothesis is indistinguishable in the log scale from the one for the
 $ \nu_{\mu} \rightarrow \nu_{\tau} $ oscillations hypothesis. 
The open circles indicate models excluded
by direct measurements and 
assume a local dark matter density of $ 0.3~ GeV~ cm^{-3} $. The DAMA indication
is at $m_x \simeq 50 - 60~ GeV$.}
\label{fig:upward}
\vspace{-0.8cm}
\end{figure}

Weakly Interactive Massive Particles (WIMPs) could be part of the galactic
 dark matter. WIMPs should be neutral particles which may form a dissipationless
gas trapped in the gravitational field of our Galaxy. Suitable WIMP candidates
should have lifetimes comparable to the age of the Universe. In SUSY models, 
like the  MSSM and SUGRA, they may be identified with the lightest neutralino;
 it is ok if R parity is violated provided it leads to a long lifetime 
neutralino. WIMPs have been searched for by direct and indirect methods.\\ 
- {\it Direct searches}. WIMPs may be searched for via their interactions in 
refined low energy detectors of $10 - 100~ kg$ mass.
 The WIMPs scatter elastically with the nuclei of
 the detector, with cross sections of the order of the weak
ones or smaller. A scattering leads to a recoil of few keV energy. The 
detectors must have low radioactivity, be well shielded and use electronics
 which reduces unwanted noise signals. The DAMA collaboration presented
results obtained with a $100~ kg$ NaI (Tl) detector,
 looking for a signal modulated
 over a one year period. They find a probable signal which could correspond
 to a neutralino
 mass of $50 - 60~GeV$ \cite{dama}.\\
- {\it Indirect methods}.
WIMPs could be intercepted by celestial bodies, slowed down and trapped in their
 centres. WIMPs and anti-WIMPs could annihilate and yield  neutrinos of GeV 
or TeV energies. The neutrinos would travel and interact below the detector
yielding high energy muons which can be detected. The search should be performed
  in small
 angular windows around
the directions of the celestial bodies. The $90\%$ C.L. 
MACRO limit for the flux from the Earth centre is $\sim 10^{-14} cm^{-2} s^{-1}$
 for a $10^{\circ}$ cone around the vertical, see Fig. 
\ref{fig:upward} \cite{ambrosio}. For the
same cone searched for around the Sun direction, the limit stands at $\sim 1.4
\times 10^{-14} cm^{-2} s^{-1}$. 

\section{HIGH ENERGY COSMIC RAYS}
\begin{figure}[htb]
\vspace{-1.8cm}
\begin{center}
\mbox{
\hspace{-2.8cm}
\epsfig{file=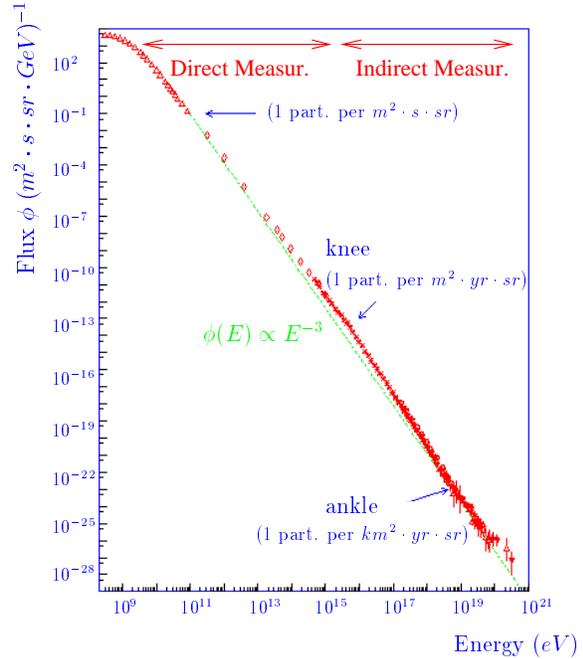,height=18.3cm}}
\end{center}
\vspace{-9cm}
\caption{\small All particle energy spectrum of primary Cosmic Rays.}
\label{fig:particle}
\vspace{-0.3cm}
\end{figure}

The all-particle spectrum of cosmic rays is shown in Fig. \ref{fig:particle} 
\cite{sorel}

\subsection{Underground muons}

Underground experiments detect a sizeable downward flux 
of high-energy 
muons, single and multiple, coming from high-energy cosmic rays. Muons 
reaching the Gran Sasso detectors traverse a minimum path length of 3100 
m.w.e. and an average one of 3700 m.w.e.. A muon must therefore have an energy 
larger than $1.3~ TeV$ to reach the detectors. 
The muon distribution in local coordinates (azimuth $\varphi$ and zenith 
$\theta$) reflects the shape of the mountain: it may be considered an 
x--ray photograph of the mountain. 
Experiments proved that the arrival time distribution of underground 
muons is random.\par

The vertical muon flux $I(h)$, where $h$ is the slant depth, was measured with 
increasing accuracy by several experiments. The flux 
 may be  
 represented by $$I(h)= B\biggl ({h_1/h}\biggr )^2 e^{-h/h_1},~~~~~~~~~~~~~~~~~~ (3)
 \eqno$$
\ndt with $B=(1.81 \pm 0.06)\times 10^{-6}~\mbox{cm}^{-2}\mbox{s}^{-1}
\mbox{sr}^{-1}$,~~ $h_1=(1231\pm 
1)\mbox{hg~cm}^{-2}$.
The muon surface flux, obtained from the measured underground muon flux, is 
$ dN_{\mu}/dE d\Omega=AE^{-\gamma}$ with $\gamma\simeq 2.78$.

\ndt{\bf Seasonal variations.} Selected muon data from several experiments 
were used to search for seasonal variations. The muon 
rate shows clear variations of about $\pm 1.4\%$ amplitude which  
repeats over the years. The muons come from pion and kaon decays 
in the upper atmosphere (at depths of less than $200~ mbar$); their intensity 
becomes greater when the atmosphere is warmer. A new measurement performed
 by AMANDA at the South pole exhibits a much larger effect ($\sim \pm 15 \%$)
reflecting the 6 month darkness and 6 month light \cite{telescopes}.
 From underground muons 
it is possible to measure the  
effective temperature of the higher atmosphere to about 
$1^{\circ}$C!\par
\ndt{\bf Muon astronomy.}
In \lq\lq muon 
astronomy" one assumes that high--energy muons remember the arrival direction of 
the parent high--energy particle with the hope that the parent particle
 has not 
deviated. Thus a search may be made for celestial 
point sources, d.c., periodic or episodic.  
The interest in muon astronomy started in 1985 with  reports  
of an excess of underground muons from the 
direction of Cyg. X--3 and with the  
Cyg. X--3 periodicity. Some reports of 
muon excesses could be connected with intense radio flares.
In order to exclude with certainty a 
variation of the muon flux from the direction of Cyg. X--3 one has to analize 
data over a long period of time.
Upper limits for a d.c. signal were established for specific sources,  
Cyg. X--3, 
Her X1, 1E2259+59, and the Crab. The d.c. limits range from 3 to $6\times 
10^{-13}~cm^{-2}~s^{-1}$ \cite{ambrosio}.
For Cyg. X--3, MACRO searched for a muon signal modulated by the 4.8 h X--ray 
period. The phase diagram does not exhibit any excess above background in any 
phase bin.
 The upper limit on a modulated sygnal is ~$F_{mod}\leq 3\times 
10^{-13}~cm^{-2}~s^{-1}$.\par
\ndt {\bf Multiple muons.} Multiple muons carry infomation about the energy 
spectrum and the chemical composition of primary cosmic rays with energies 
 $\geq 50$ TeV. The sensitivity to composition arises from the fact 
that heavy nuclei are more effective than protons in producing multiple 
muons. The  measurable distributions are:
i) The decoherence function (the distribution of the distance 
between two muons) \cite{sioli}.
ii) The decorrelation function (the double--differential
distribution of two muon 
relative angles).
iii) The multiplicity distribution.
iv) The muon group sub-structure.

\begin{figure} [htb]
\vspace{-0.2cm}
\begin{center}
\mbox{
\hspace{-0.3cm}
\epsfig{file=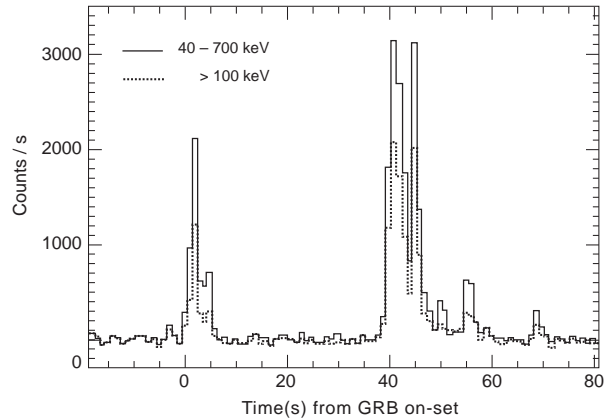,height=5.5cm}}
\end{center}
\vspace{-1cm}
\caption{\small Examples of $\gamma$-ray bursts at x-ray energies (BeppoSAX).}
\label{fig:ray}
\vspace{-0.8cm}
\end{figure}

The analyses require a model of the hadronic interactions at high energy 
(nucleon--nucleus and nucleus--nucleus), trial models of the energy variation of 
the composition of cosmic rays,  simulation of the cascade in air and in the 
rock, and a simulation of the detector. In practice  
one uses iteration procedures with continuous improvements in 
models and simulations, and eventually a 
  multiparameter fit of all avaible data.\par
A slow increase of the average primary mass is observed when going 
from $10^3$ to $10^4~ TeV$, i.e. when crossing the \lq\lq knee" 
 of the cosmic ray all--particle flux \cite{sioli} \cite{scapparone}.

It has to be noted that the muons in the same bundle arrive at the same time to
within few $ns$.

\subsection{Cosmic rays of highest energies}

The origin of high--energy cosmic rays is essentially unknown, and it is difficult 
to devise acceleration mechanisms for the highest energy cosmic rays. 
Recently magnetic monopoles of relatively low mass accelerated by the 
galactic magnetic field to high
 energies and high velocites have been proposed as possible sources of the
 highest energy cosmic rays.
It should be remembered that cosmic ray nuclei with energies  $> 4\times 
10^{19} eV$ cannot come from distances $> 50~ Mpc$ because of the 
Greisen cut--off caused by  the interaction of protons with the $2.7 K$
 photons of 
the cosmic backround radiation (at these energies the c.m. 
$p\gamma$ energy is above pion threshold, the cross section becomes large, 
 and cosmic rays are soon degraded in energy).\par

It is clear that more data are needed and that this requires large
Extensive Air Shower Arrays (EAS). We have heard reports from KASKADE 
and EASTOP \cite{scapparone}. The
largest new project is AUGER, with a first array in South America;
 a similar array will also be built 
 in North America. Each Auger array will cover $\sim 5000~ km^2$, with
 different types of
 detectors (hybrid air shower detectors): sampling  water tanks and
 improved fly's Eyes detectors which detect the nitrogen luminescence in
 the atmosphere, thus
measuring the shower profile \cite{zavrtanik}.

Among the different types of proposed detectors for large arrays we  have heard
Sorel's presentation about the possibility of using standard solar pannels
connected in series/parallel to detect the \^Cerenkov light \cite{sorel}.

\section{SCIENTIFIC EXPLORATION OF SPACE}
At this meeting we had a number of reports on physics and astrophysics
research performed with balloons, satellites and the space station. There is
an increasing effort in this field.

Recently balloon experiments have been performed to measure cosmic ray muons in
 the atmosphere \cite{circella}: these measurements are relevant for a more precise
 determination of the atmospheric neutrino flux.

The BeppoSAX satellite  measured x-ray bursts,  identifying 14 x-ray 
sources as the counterparts of $\gamma$-ray bursts (see Section 6) 
\cite{fontera}.

The AMS (Alpha Magnetic Spectrometer) experiment on the  space 
station should make a thorough search for antimatter, measure the cosmic ray
composition, and perform other searches. 
A test flight was successful and it has already provided important 
information on the flux of $p, d,~ ^3He,~ ^4He$, and limits on antimatter 
\cite{alpat}.

A variety of experiments are becoming realities. For instance PAMELA is 
measuring $\ov p, e^+$ of 100-200 GeV and is making a search for $\ov {H e}$
nuclei \cite{spartoli}.
AGILE should be operative in 2002 \cite{spartoli}; it has several $\gamma$ ray detectors 
optimized to cover different $\gamma$ ray energies above $30~ MeV$,
preliminary results have been obtained by NINA, etc. \cite{spartoli}.

\section{ $\gamma$ -RAY BURSTS}
Since few years, the  observation of $\gamma$-ray bursts (GRBs) poses one of 
the main misteries of astrophysics. The $\gamma$-ray observatory, on 
board the Compton satellite, observed every day, a new $\gamma$-ray burst of MeV
 energy. The burst durations are from 30 ms
to 1000s (but this depends on the sensitivity and time resolution of the 
experiment: one may only see the tip of an iceberg).  
The bursts come from all directions of space; in almost all cases they 
represent a single episode. The rise time of the bursts is very fast, and this
suggests that they could be connected with neutron stars. Measurements from
the BeppoSAX satellite observe the GRBs at x-ray energies, Fig. \ref{fig:ray},
 determining
more accurately the position of the source; they see a tail in intensity 
(afterglow). When seeing a burst, BeppoSAX alerts 
the astronomical community; it was thus possible to observe the optical
counterparts of x-ray emitters; the optical signal lasts  a few days
 \cite{fontera}; at least one appeared to be at 
the border of a far away galaxy. It should be stressed that, even if it is
 seen in x-ray and in the visible, most of the emitted energy of GRBs is in
 $\gamma$ rays, at MeV energies.
   
It would be interesting to observe the GRBs at higher energies (multi GeV)
\cite{iacovacci}. It would even be more interesting to observe them with
 neutrinos; trials are being made, but probably one needs larger 
neutrino telescopes, with lower energy thresholds.

\section{RARE PARTICLES AND PROTON DECAY SEARCHES}
\subsection{Magnetic Monopoles}

Grand Unified Theories (GUTs) of electroweak and strong interactions predict
the existence of magnetic monopoles (MMs) 
with large mass, larger than  $10^{16}$~GeV, and
magnetic charges $g = ng_{\tiny Dirac} = nc/2e = n68e$, with $n=1,~2,...$ . 
These
theories leave  open the question of monopole abundance. MMs were probably  
produced in the early universe, at the end of the GUT era as point defects; others may have been 
produced in ultra--high energy collisions. Standard
cosmology predicts too  many monopoles, whereas models with inflation at the
GUT phase transition predict very few. Several superstring models predict the
existence of multiply--charged MMs $(n=3)$.
In some models, the primordial monopoles appeared when the temperature
of the universe reached relatively low values. These monopoles were probably 
not  diluted by inflation in the early universe. The existence of
large--scale magnetic fields, on the galactic scale, leads to an astrophysical
constraint, the so--called Parker bound, with an upper limit on the monopole
flux at the level of $10^{-15}~cm^{-2}~s^{-1}~sr^{-1}$;
 an extended Parker bound leads to a flux limit almost an order of magnitude
 smaller.

\begin{figure}[htb]
\vspace{-0.8cm}
\begin{center}
\mbox{
\epsfig{file=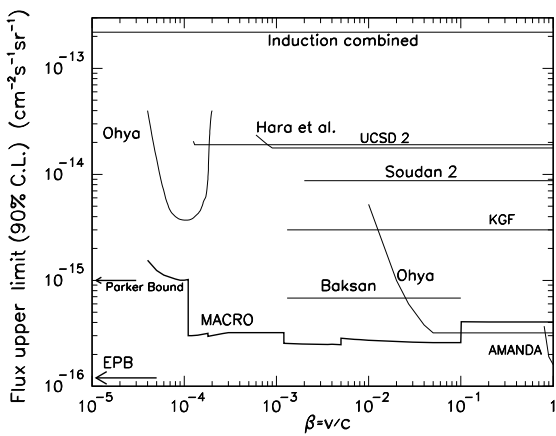,height=6.3cm}}
\end{center}
\vspace{-1cm}
\caption{\small Compilation of 90 \% C.L. upper limits for a flux of heavy
magnetic monopoles [32].}
\label{fig:neutral}
\vspace{-0.8cm}
\end{figure}

Underground experiments have searched for 
MMs  in the penetrating cosmic radiation 
 using scintillators, gas tubes, and nuclear track detectors via
 $dE/dx$, time of flight and pulse shape 
analyses. At present there are only a few large  experiments. They  
tested the sensitivity of their detectors to low velocity MMs. 

 New limits have
 been presented by MACRO \cite{patrizii} and by AMANDA \cite{telescopes}.  
The present limits on massive cosmic MMs are summarized in
 Fig. \ref{fig:neutral} for 
$g=g_D$ bare poles with mass $> 10^{16}$ GeV and for catalysis cross 
sections smaller than few  mb.

\subsection{Dark Matter}

Analyses of the rotation curves of stars in galaxies, and of galaxies in 
clusters of galaxies prove (assuming the validity of Newton's law) that most of 
the matter is unseen:
~~$ \Omega_{vis}\simeq 0.01~~,~~\Omega_{halo}\simeq 0.1~~,~~
\Omega_{TH}\simeq 1,~~\hbox{where}~\Omega=\rho/\rho_c$.
\par
The unseen DM could be: a) baryonic, in the form of gas, planets 
like jupiter, brown dwarfs, nuclearites; b) non baryonic, i.e. a gas of 
particles. In the latter case there could be: i) hot DM, i. e. particles which 
were relativistic when in the early universe they decoupled from the rest of 
matter and radiation (an example could be neutrinos with a mass of a few eV); 
ii) cold DM, i.e.  particles which were non--relativistic at decoupling (for 
example 
 the WIMPs, see Section 3.2).
 These particles are probably  located in the galactic halos; their abundance in
 the vicinity of
the solar system could typically be $\sim~0.3~ GeV/cm^3$, and their velocity 
 $\sim300~ km/s$.

\ndt {\bf Nuclearites.}
 The hypothesized stable phase of quark matter, called  
strange quark matter or
 nuclearites, formed by quarks $u,~d$ and $s$, may be the true ground state of 
QCD. Nuclearites could have 
masses ranging from a  few GeV to the mass of a neutron star. Because of
this wide range,  searches were performed using  a variety of experimental 
techniques.
At this meeting new results have been presented,
 using techniques developed for MM searches. Limits 
 for nuclearites with masses larger than
 $0.1~ g$ (which can penetrate the earth) are at the level of the 
limits of MMs \cite{patrizii}; the limits are twice as large 
for nuclearites with m $< 0.1~ g$, which cannot traverse the earth.

 Other 
"exotic" objects, like the Q-balls (aggregates of squarks, sleptons and 
Higgs fields) have also been discussed at this meeting \cite{patrizii}. 

\subsection{Proton decay}

GUTs place quarks and leptons in the same multiplets. Quark 
$\longleftrightarrow$ lepton transitions are thus possible and should be mediated 
by supermassive vector bosons $X,Y$ with $m\sim 10^{14}$ $GeV$. A free proton 
 may decay as $N\to\ell^++$ 
meson(s) or $N\to\ov\nu+$ meson(s). Proton decay, with a predicted lifetime 
of the order of $10^{31}~ y$, motivated the construction of the first 
underground detectors 
with masses of the order of $1000~ t$. Present  detectors are either water 
\^Cerenkov detectors (SK) or  tracking 
calorimeters (Soudan 2). Water detectors have larger masses and more free 
protons and may detect the sense of the track direction. 
Tracking calorimeters have a higher spatial resolution and a better $\pi/\mu$ 
separation at energies of about $200~ MeV$. Technical developments are being made 
towards a TPC type liquid chamber (ICARUS). A $3~ t$ prototype works 
well and a $600~t$ module is under construction.
Superkamiokande presented the following limits

$
\left\{
\begin{array}{ll}
\tau BR(p\to e^+\pi^{\circ})&>1.6 \times 10^{33} ~~y\\
\tau BR(p\to \overline\nu K^+)&>6.7\times 10^{32} ~~y
\end{array}
\right.
$\par

\ndt at $90\%$ C.L. \cite{shiozawa};
 they rule out the simplest SU(5) GUT models.

\section{SELECTED RESULTS FROM ACCELERATOR EXPERIMENTS}

\begin{figure}[htb]
\vspace{0cm}
\begin{center}
\mbox{
\epsfig{file=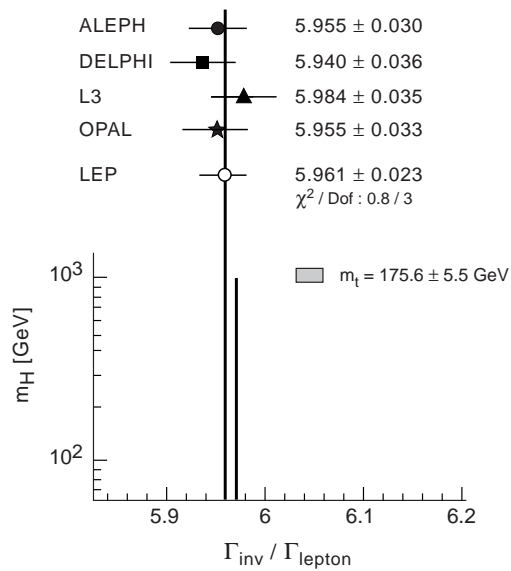,height=7.5cm}}
\end{center}
\vspace{-1cm}
\caption{\small Ratio of the invisible width relative to the leptonic width
at LEP. From this type of data, using updates, one obtains now that the number
 of neutrino families is
2.9835 $\pm$ 0.0083.}
\label{fig:width}
\vspace{-0.8cm}
\end{figure}

{\bf LEP.} The four experiments at the LEP positron-electron collider provided
new improved precision values of the $Z^\circ$ lines shape. The $Z^\circ$ mass
is now known with a precision of two parts in $10^5$, and has acquired the 
status of one of the three basic inputs of the Standard Model (SM) of particle
physics. An important quantity derived from the line shape parameters is the 
number of light neutrino species which is now\\ 
$N_\nu= \left( \frac{\Gamma_{inv}}{\Gamma_l} \right) /
\left( \frac{\Gamma_{\nu}}{\Gamma_l} \right)_{SM} = 2.9835 \pm 0.0083$,  
 see Fig. \ref{fig:width}; a direct method (the neutrino counting method) 
confirms this result \cite{lep}.

\begin{figure}[htb]
\vspace{0cm}
\begin{center}
\mbox{
\hspace{-0.3cm}
\epsfig{file=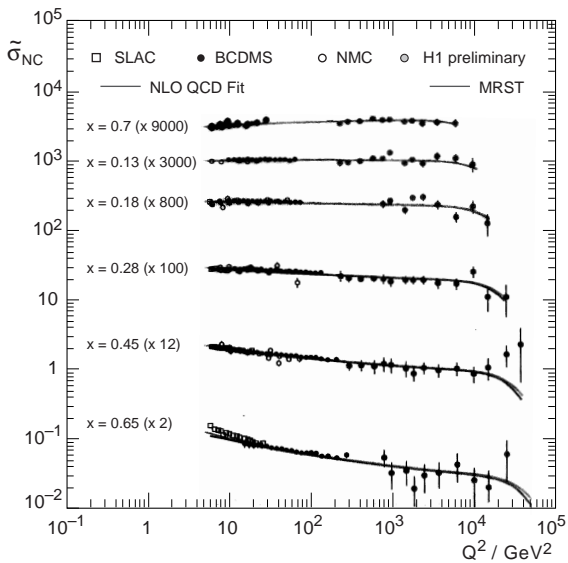,height=7.5cm}}
\end{center}
\vspace{-1cm}
\caption{\small Neutral current reduced cross sections vs $Q^2$ measured  in 
$e^+ p$ collisions at  HERA.}
\label{fig:current} 
\vspace{-0.8cm}
\end{figure}

 From this determination one may deduce the
 amount of helium expected in primordial nucleosynthesis: one expects $24 \%$,
in fair agreement with astrophysical data. The charged lepton universality
is now established at the $0.1 \%$ level; the muon and the $tau$ lepton appear
 more and more to be replicas of the electron. The increased energy of LEP 
(LEP2) allowed to study the reaction $e^+ e^- \rightarrow Z^\circ \rightarrow
 W^+ W^-$, proving the existence of the triple boson vertex, $Z^\circ
 W^+ W^-$, and a precise measurement of the $W^\pm$ mass, which very likely
 will become one of the three inputs of the SM. LEP1 allowed a detailed study
of QCD properties, in particular  a precise determination of the strong 
coupling constant and of its variation with energy (LEP2 is showing that the
 variation continues to higher energies). It may be worth remembering that 
precision measurements lead to the first determination, below threshold, of 
the mass of the quark top and now gives a hint of the mass of the
 Higgs boson. LEP2 with data
up to $\sqrt s = 202~ GeV$ yields direct limits on the S.M. Higgs boson,
 $m_{H^\circ} > 103~ GeV$, and
 on a variety of particles prediced by models beyond the SM.\\
{\bf HERA.} At the asymmetric $e^+ p$ collider ($E_e = 26.7~ GeV,~ Ep = 820~ 
GeV$)
at Hamburg, two experiments are providing a wealth of information on CC deep
inelastic scattering, in particular at very small values of $x$ and large
values of $Q^2$. They also measure the neutral current (NC) cross section, see
Fig. \ref{fig:current} \cite{corradi}; notice that it is
 related to parton densities and takes
 into account effects of $x F_3$. A considerable part of the HERA program 
concerns the searches for particles predicted by models beyond the SM 
\cite{corradi}. In particular new more stringent limits on leptoquarks have 
been presented.   
 
\vspace{-0.1cm}

\section{GRAVITATIONAL WAVES}

The earth should be continously bombarded by gravitational waves produced by
distant celestial bodies subject to "strong" gravitational effects. The 
amplitude of the gravitational wave emitted by a celestial body is
 proportional to its mass, to its acceleration, and to the inhomogeneity in its
 mass distribution. Gravitational waves are emitted when the quadrupole moment
of an object of large mass is subject to large and fast variations. Only
large celestial bodies subject to unusual accelerations should produce sizeable
gravitational radiation measurable on earth. These bodies may be binary systems
of close-by stars (in  particular when a neutron star is about to fall on
 the other); they yield a periodic emission of gravitational waves, with 
frequencies from few hundred Hz to 1 MHz. Asymmetric supernovae explosions
 may give
 bursts of gravitational waves, with frequencies of the order of 1 kHz over few
ms. Also vibrating black holes, star accretions, galaxy formation, and the 
Big Bang may produce or have produced gravitational waves.

A gravitational wave is a transverse wave which travels at the speed of light.
A gravitational wave should modify the distances between objects in the
 plane perpendicular to the direction of propagation of the wave. These 
deformations are expected to be extremely small. It has been estimated that a
star collapse  at the centre of our galaxy may produce a variation of the 
order of $h \sim 10^{-18}$ metre per metre of separation of two objects on 
earth.  The Supernova 1987A in the large Magellanic Cloud could probably have
 produced 
a distortion 10 times smaller. A collapse in the Virgo cluster (at MPc), should
yield relative variations of $10^{-21}$.

Very sensitive instruments are needed to observe gravitational waves. The two
 main lines developed until now are resonating bars at low temperatures and
 long laser interferometric systems. A major program is underway for both types 
of detectors, hoping to be able to detect gravitational star collapses up to 
the Virgo cluster, corresponding to $h \sim 10^{-21}$. The supercooled (0.1 K)
bars NAUTILUS and AURIGA are operating in Frascati and Legnaro (Padova), 
respectively. A pair of long interefometers (LIGO) are under construction in
 the US, while one long interferometer is under construction at Pisa (VIRGO)
\cite{virgo}.
Several detectors, in coincidence, are needed to ensure  that the observed 
signal 
is not spurious.

The detection of gravitational waves would have far reaching consequences.
 It would prove the validity of the general theory of relativity; in 
astrophysics it would open up a new observational window related to violent 
phenomena in the universe.

\section{CONCLUSIONS}

We had an interesting and lively workshop on topical subjects. Many new
 interesting results were presented, as well as many new proposals: the field of
astroparticle physics in general and of neutrinos in particular is very alive
 \cite{proceedings} and we look forward to many new exciting results in the
 future.

I would like to thank Ms. Luisa De Angelis for typing the manuscript. I 
acknowledge the cooperation of all the colleagues at the workshop and the help
of the colleagues in Bologna, in particular of Dr. M. Giorgini.


\vskip -0.3cm 


\begin{thebibliography}{9}

\bibitem{soudan} Soudan 2 Collaboration, Proc. of NEUTRINO 98, Takayama, 
Japan; 
W.W.M. Allison  {\it et al.}, Phys. Lett.  B449 (1999)137.

\bibitem{ronga98} MACRO Collaboration, Proc. of NEUTRINO 98,
  hep-ex/9810008; M. Ambrosio {\it et al.}
Phys. Lett. B434 (1998)451; hep-ex9904; INFN/AE-99/09, INFN/AE-99/10; M.
 Spurio, this workshop. 

\bibitem{sk98} SuperKamiokande Collaboration, 
Proc. of NEUTRINO 98, hep-ex/9810001; Y. Fukuda  {\it et al.}, Phys. Rev.
 Lett. 81 (1998) 1562.

\bibitem{kajita} T Kajita, this workshop.

\bibitem{agrawal} V. Agrawal {\it et al.}, Phys. Rev.  D53 (1996) 1314.
                                  
\bibitem{gluck} M. Gl\"{u}ck {\it et al.},
Z. Phys. C67 (1994) 433.

\bibitem{lohmann} W. Lohmann {\it et al.}, CERN-EP/85-03 (1985).

\bibitem{fogli} G. Fogli, S.F. King, D.P. Roy, W.G. Scott, this workshop.

\bibitem{battistoni} G. Battistoni, this workshop.

\bibitem{circella} M. Circella, this workshop.

\bibitem{terranova} F. Terranova, this workshop.

\bibitem{bellini} G. Bellini, A. Grandpierre, this workshop.

\bibitem{stancu} I. Stancu, this workshop.

\bibitem{kawanura} T. Kawamura, this workshop.

\bibitem{harris} P. Harris, this workshop.

\bibitem{kamland} G. Gratta (KAMLAND), this workshop.

\bibitem{kignoli} C. Vignoli, this workshop.

\bibitem{cocco} A. Cocco, this workshop.

\bibitem{weinheimer} C. Weinheimer, this workshop.

\bibitem{fiorini} E. Fiorini, Proc. of the Neutrino Telescope Workshop, Venezia 
(1999) page 35.

\bibitem{ambrosio} M. Ambrosio, {\it et al.}, INFN/AE-99/11; INFN/AE-99/12, 
INFN/AE-99/13.

\bibitem{telescopes} Neutrino telescopes: S. Bottai, C. Carlogam, P. Niessen, this workshop.

\bibitem{dama} DAMA Collaboration, R. Barnabei {\it et al.}, Phys. Lett. B450
 (1999)
488.

\bibitem{sorel} M. Sorel, this workshop.

\bibitem{sioli} M. Sioli, E. Scapparone, this workshop.

\bibitem{scapparone} K. Bernl$\ddot{o}$hr, A. Chiavassa, this workshop.

\bibitem{zavrtanik} D. Zavrtanik, this workshop.

\bibitem{fontera} F. Frontera, A. Iyudin, this workshop.

\bibitem{alpat} A. Alpat, this workshop.

\bibitem{spartoli}A. Morselli, R. Sparvoli, P. Spillantini, this workshop.

\bibitem{iacovacci} M. Iacovacci, this workshop.

\bibitem{patrizii} L. Patrizii, M. Giorgini, M. Ouchrif, R. Wigmans, 
this workshop.

\bibitem{shiozawa} SuperKamiokande Collab., M. Shiozawa {\it et al.},
 Phys. Rev. Lett. 81 (1998)
3319; Y. Hayats {\it et al.}, ICRR-453-99-11 (1999).
                  
\bibitem{lep} The LEP Collaborations, CERN-EP/99-15; M. Chemarin, this workshop.

\bibitem{corradi} M. Corradi, C. Beier, this workshop.

\bibitem{virgo} VIRGO, R. Passaquieti, this workshop.

\bibitem{proceedings} Proceedings of the $5^{th}$ School on Non Accelerator 
Particle Astrophysics, Trieste University Press (1999). 

\end{thebibliography}
\end{document}